\documentclass{article}

\usepackage{microtype}
\usepackage{graphicx}
\usepackage{subfigure}
\usepackage{booktabs} 
\usepackage{tabularx} 

\usepackage{multirow} 
\usepackage{amsmath}
\usepackage{svg}

\usepackage{hyperref}

\usepackage[preprint]{mlsys2023}

\mlsystitlerunning{RTP: Rethinking Tensor Parallelism with Memory Deduplication }

\begin{document}

\twocolumn[
\mlsystitle{RTP: Rethinking Tensor Parallelism with Memory Deduplication }



\mlsyssetsymbol{equal}{*}

\begin{mlsysauthorlist}
\mlsysauthor{Cheng Luo}{Independent}
\mlsysauthor{Tianle Zhong}{Virginia}
\mlsysauthor{Geoffrey Fox}{Virginia}
\end{mlsysauthorlist}

\mlsysaffiliation{Independent}{Independent Researcher}
\mlsysaffiliation{Virginia}{Computer Science Department and Biocomplexity Institute, University of Virginia, Charlottesville, VA, USA}

\mlsyscorrespondingauthor{Cheng Luo}{wdlctc@gmail.com}

\mlsyskeywords{Machine Learning, MLSys}

\vskip 0.3in

\begin{abstract}
In the evolving landscape of neural network models, one prominent challenge stand out: the significant memory overheads associated with training expansive models. 
Addressing this challenge, this study delves deep into the Rotated Tensor Parallelism (RTP). RTP is an innovative approach that strategically focuses on memory deduplication in distributed training environments. It boasts of unique features like a customized communication primitive and the Flyweight Pattern initialization. Furthermore, RTP ensures a seamless overlap between partition computation and partition weight communication, optimizing the training process. Our empirical evaluations underscore RTP's efficiency, revealing that its memory consumption during distributed system training is remarkably close to the optimal - distributing the memory overhead of a single machine equitably among multiple machines. The experimental results demonstrate that RTP is capable of achieving comparable performance to Distributed Data Parallel while providing support for significantly larger models with near-linear scalability in terms of memory. Code of RTP is available at \href{https://github.com/wdlctc/rtp}{https://github.com/wdlctc/rtp}.

\end{abstract}
]



\printAffiliationsAndNotice{}  

\section{Introduction}
\label{submission}

The landscape of machine learning has been dramatically reshaped by the exponential growth of neural network models. These architectures, with their intricate layers and vast parameter sets, have catalyzed breakthroughs across diverse domains.
As these neural behemoths continue to push technological boundaries, the challenges they pose become increasingly multifaceted. Beyond the complexities of their design and deployment lies the colossal task of training them. Given the massive storage overhead that comes with its scale, there is a growing need for industry-grade tools that can effectively simplify and accelerate the inference and training process. These tools often face scalability and efficiency issues in distributed environments.

To better understand the theoretical computation and memory consumption, we invoke the theoretical underpinnings of Unlimited Memory Idealized Computer stands as an epitome of unbounded potential. With its infinite memory like Turing machine  \cite{vzak1983turing} , it represents an idealized scenario where any computational problem, irrespective of its scale, can be addressed without memory constraints. When juxtaposed against the realm of neural network training, the Unlimited Memory Idealized Computer serves as an optimal benchmark. In this ideal world, a neural network model, no matter how vast, would fit perfectly within the memory confines of, consuming memory in the most optimal manner.

\begin{table*}
    \centering
    \begin{tabular}{c|c|c|c}
        Techniques & Activations Memory Cost & Parameters Memory Cost&  Memory Duplication \\
        \hline
        \multirow{2}{*}{No parallelism} & \multirow{2}{*}{$A$} & \multirow{2}{*}{$W + G$} & \multirow{2}{*}{0} \\ & & & \\ \hline
        \multirow{2}{*}{Tensor parallel} & \multirow{2}{*}{$A \times N$} & \multirow{2}{*}{$W + G$} & \multirow{2}{*}{$A \times (N-1) $}  \\ & & & \\ \hline
        \multirow{2}{*}{Data Parallel} & \multirow{2}{*}{$A$}  & \multirow{2}{*}{$(W + G) \times N $} & \multirow{2}{*}{$(W + G) \times (N - 1) $} \\ & & & \\ \hline\multirow{2}{*}{Pipeline Parallel} & \multirow{2}{*}{$A + A_p \times N$}  & \multirow{2}{*}{$W + G$} & \multirow{2}{*}{$A_p \times N$}  \\ & & & \\ \hline
        \multirow{2}{*}{FSDP} & \multirow{2}{*}{$A$}  & \multirow{2}{*}{$W + G + max(W, G) \times (N-1)$} &  \multirow{2}{*}{$max(W, G) \times (N-1)$} \\ & & & \\ \hline
        \multirow{2}{*}{RTP} & \multirow{2}{*}{$A$}  & \multirow{2}{*}{$W + G + max(W, G)$} & \multirow{2}{*}{$max(W, G)$}  \\ & & & \\ \hline
        \multirow{2}{*}{RTP Inplace} & \multirow{2}{*}{$A$}  & \multirow{2}{*}{$W + G $} & \multirow{2}{*}{$0*$}  \\ & & & \\ \hline
    \end{tabular}
    \caption{Activations memory,  Params Memory Duplicated Memory Buffer for total distributed system using different techniques. $N$ refers to number of workers, $A, W, G$ refer the activation, weight and gradient memory, $A_p$ is the intermediate activation of pipeline layer}
    \vspace{-0.15in}
    \label{tab:memory-eqns}
\end{table*}

However, as we transition from this theoretical abstraction to the tangible challenges of real-world parallel training \cite{narayanan2019pipedream} and inference \cite{pope2023efficiently}, discrepancies emerge. The memory consumption of practical parallelism techniques often exceeds the optimal memory footprint represented by the idealized computer. This excess can be conceptualized as "Memory Duplication." Here we dissect various parallelism strategies elucidating their respective memory footprints and the consequent memory duplication:

\begin{itemize}
\item {\bf Tensor Parallelism}: Tensor parallelism \cite{shoeybi2019megatron} shards the model's weights across multiple devices, allowing for concurrent computation of large tensors. However, the batch inputs and activations are copied across multiple machines, leading to memory duplication.

\item {\bf Data Parallelism}: This technique \cite{li2020pytorch} distributes different data subsets across devices. While it shards the batch inputs and activations, each machine holds a complete copy of the model's weights and gradients, introducing significant memory duplication.

\item {\bf Fully Sharded Data Parallelism (FSDP)}: FSDP \cite{zhao2023pytorch} aims to reduce memory duplication by temporarily sharding the model's weights and gradients and reconstructing them on-demand. However, the time taken for reconstruction still introduces memory duplication. 

\item {\bf Pipeline Parallelism}: By segmenting the model into stages processed on different devices, pipeline parallelism \cite{narayanan2019pipedream} can reduce the memory footprint. However, the need to store intermediate activations on multiple devices introduces memory duplication.

\item {\bf Zero-Redundancy Parallelism}: \cite{ren2021zero} seeks to move the memory duplication from the GPU to the CPU, aiming to ensure each model parameter or activation is stored just once across all devices. While it promises optimal memory usage, the shift from GPU to CPU can introduce its own set of challenges.
\end{itemize}

In this paper, we delve deeper into the memory intricacies of distributed training, focusing on a dimension that has been largely overlooked: memory deduplication \cite{bugnion1997disco}. We introduce RTP (Rotated Tensor Parallelism), a novel technique that seeks to minimize memory duplication by strategically sharding activations and rotating weights/gradients.

Our comprehensive analysis, detailed in Table \ref{tab:memory-eqns} , juxtaposes the memory requisites of various distributed training methodologies, underscoring the ubiquity of memory duplication. We postulate that an optimal training framework necessitates a paradigm wherein GPUs retain only sharded activations and parameters, thereby minimizing memory overheads. Conventional data parallelism, albeit efficacious in certain scenarios, inherently mandates that GPUs maintain a complete layer for both forward and backward propagation, an attribute that renders it suboptimal for achieving desireable memory efficiency.

Our discussion leads us to the introduction of RTP, a fresh perspective on tensor parallelism. At the heart of RTP's design is the integration of the "rotation primitive" combined with the Flyweight Memory Pattern. This unique combination allows RTP to shard both model parameters and activations efficiently, and prefetch weights from neighbor worker. The primary goals are twofold: a significant reduction in memory usage and a boost in training performance. Furthermore, RTP's rotation primitive ensures a seamless overlap with computation, eliminating idle GPU time. Through this mechanism, RTP achieves a harmonization of FLOPS and memory usage, closely mirroring the optimal benchmarks established by the idealized computer. 

Preliminary evaluations indicate that RTP not only holds its ground against established methodologies like FSDP but also exhibits pronounced advantages. Notably, RTP has demonstrated potential memory savings in excess of 75\% compared to FSDP with comparable throughput performance.

\section{Background}

\begin{figure*}
    \centering
  \includegraphics[scale=0.35]{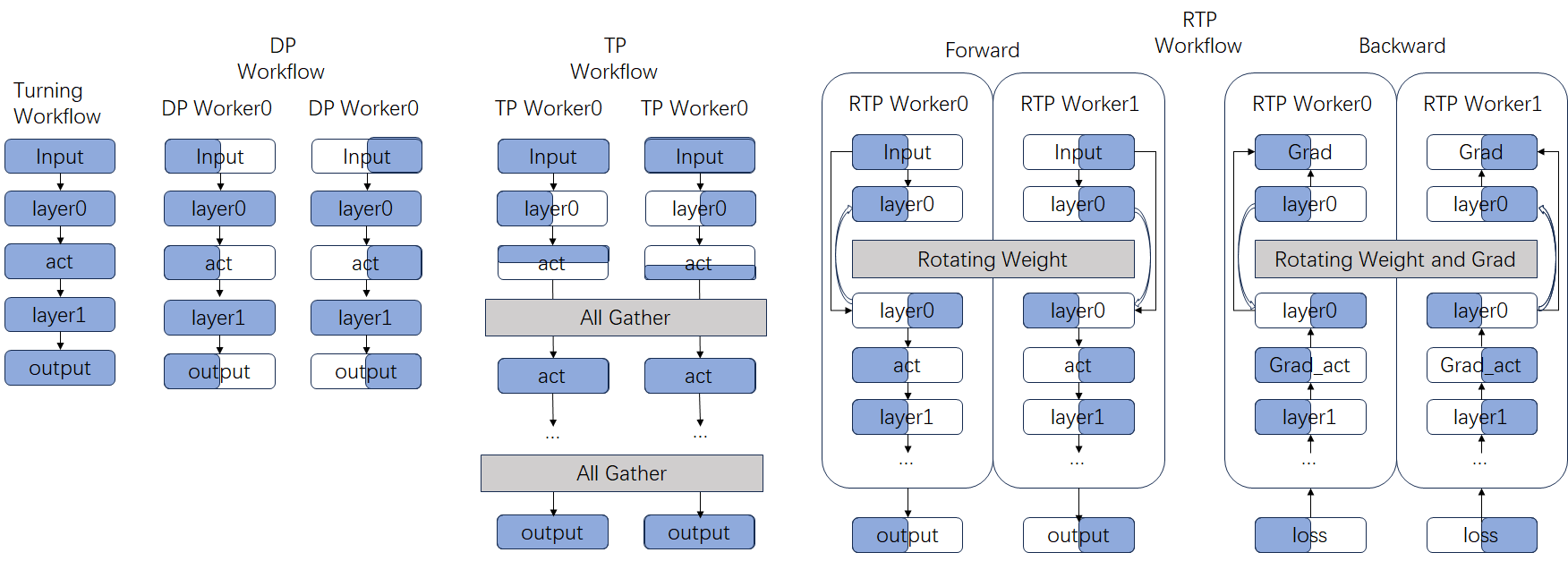}
  \caption{RTP Algorithm Overview }
  \label{fig:RTP Algorithm Overview}
  
    \vspace{-0.15in}
\end{figure*}

Large Language Models (LLMs) have become a cornerstone in the realm of machine learning with billions of parameters (llama 7B-70B\cite{touvron2023llama}, PaLM2 340B\cite{dehghani2023scaling}, GPT-4 1.76T\cite{openai2023gpt4}). The magnitude of these models can be so overwhelming that accommodating them within the memory of a singular processor becomes unfeasible (A100 80G \cite{a100_80G}). Consequently, conventional training methodologies becomes impossible when faced with LLMs.
chan
To address this, distributed frameworks emerges as a pivotal solution. At its core, distributed training/inference is about partitioning \cite{pope2023efficiently}, a strategy that disperses the training workload across multiple processors with two primary components: activations and parameters. By efficiently distributing these components across processors, distributed frameworks including Horovod\cite{sergeev2018horovod},amazon sagemaker \cite{liberty2020elastic}, Pytorch \cite{imambi2021pytorch}, Tensorflow \cite{singh2020introduction} enable the seamless and efficient training/inference of LLMs, ensuring that the potential of these vast models is fully realized without being hindered by memory constraints.

\subsection{Activation Partition}

Activation Partitioning techniques are meticulously crafted to distribute the input dataset and activations across multiple devices, optimizing memory usage and computational efficiency during training. Data parallelism \cite{li2020pytorch, dean2012large}, a cornerstone in this domain, partitions the activations across devices, each handling a distinct subset. This approach, while offering near-linear scaling efficiency and accelerating time-to-train with minimal code modifications, grapples with the challenge of model weight duplication across all devices, leading to large parameter memory duplication scaling up with the number of devices. 

\subsection{Parameter Partition}
Parameter Partitioning techniques focus on optimizing the distribution of model parameters across devices to enhance training efficiency. Tensor Parallelism \cite{shoeybi2019megatron, narayanan2021efficient}, a pivotal technique in this realm, shards model parameters to facilitate partial computations on individual devices. By communicating activations at necessary layer boundaries, it achieves a harmonious balance between memory usage and computational efficiency, streamlining the training process. However, the activation duplication are also  scaling up with the number of devices. 

\subsection{Holistic Partition}
Holistic Partitioning seeks to provide a unified approach to distributed training by synergistically optimizing both activations and parameters. Within this paradigm, several techniques have emerged that offer comprehensive solutions:

\textbf{Fully Sharded Data Parallelism (FSDP)}: FSDP \cite{zhao2023pytorch} offers a panoramic approach to model optimization. By segmenting a model into digestible units, then flattening and sharding all parameters within these units, FSDP achieves optimal memory utilization. The sharded parameters are communicated and reconstructed as needed for computations, and subsequently discarded, ensuring minimal peak memory usage. This intricate methodology, encompassing model decomposition, parameter sharding, and on-the-fly communication, guarantees both computational efficiency and memory optimization.

\textbf{Pipeline Parallelism}: Pipeline Parallelism \cite{gpipe, kim2020torchgpipe, pipetransformer, narayanan2019pipedream} introduces a methodical approach to distributed training. By dividing the model into distinct stages and distributing them across devices, it ensures that both parameters and activations are efficiently partitioned. However, this structure requires communication between stages with intermediate activation.

RTP also emerges as an all-encompassing control mechanism in distributed training, addressing both weights and activations. A standout feature of RTP is its ability to significantly reduce memory duplication, thereby enhancing its efficiency in distributed training scenarios.

\section{Design}

RTP is designed to scale and accommodate large models by leveraging the principles of sharding for both dense parameters and batched activations. By dissecting models into smaller, autonomous modules, RTP ensures that each worker can execute forward and backward operations independently. 

Distinct from its contemporaries, RTP's approach to parameter and gradient management is streamlined. While FSDP reconstructs full parameters and gradients, RTP maintains sharded parameters and gradients of only one unit at any given moment. This design choice, rooted in the capability of each sharded parameter to function independently, ensures optimal memory usage and computational efficiency.

Figure \ref{fig:RTP Algorithm Overview} offers a visual representation of RTP's workflow which adopts a rotation mechanism for sharded parameters and gradients across different workers. This example employs two workers as an exemplar. The forward pass witnesses a dynamic shift of weights between workers. Worker 1, initiating the forward pass based on its first shard of weight, subsequently transmits its weight to the next worker while simultaneously receiving weights from the preceding worker. This iterative process, executed N-1 times, culminates in Worker 1 holding the weight of the subsequent worker, rather than its original weight.
The backward pass operates in a mirror fashion. The weights traverse in reverse, ensuring that by the end of the process, each worker recovers its original weight. This design ensures that each worker is involved in the send/receive process for every sharded parameter exactly once, promoting balanced workload distribution and minimizing communication overheads.

Moreover, RTP introduces two distinct variations:
\begin{itemize}
\item {\bf In-place RTP}: This version is optimized for memory efficiency. Its memory requirements closely mirror those of a idealized computer, ensuring minimal memory overhead during training. The in-place operations \cite{In-place_Operations} ensure that the existing memory is reused, thereby reducing the need for additional allocations.

\item {\bf Out-of-place RTP}: This version is optimized for throughput. By enable overlapping and multi-stream processing \cite{sourouri2014effective}, it ensures that the computational flow is unblocked, leading to faster processing times.

\end{itemize}


\subsection{Model Initialization}

Our model initialization aims distribute various sub-modules across multiple devices with alterations to the model's initialization code. However A predominant obstacle emerged due to the intrinsic architecture of renowned deep learning frameworks such as PyTorch and TensorFlow. These platforms, by default, bind forward propagation with reverse transmission. This implies that even if parameters are interchanged between devices during forward propagation, the subsequent back propagation remains anchored to the initial weights. To realize a cohesive gradient generation-communication-update cycle, the deployment of multiple module instances without additional memory cost became significant challenge for RTP. 

To address the aforementioned challenges, RTP introduced the \emph{Flyweight Pattern initialization} inspired by \cite{harmes2008flyweight}. This innovative approach involves the creation of multiple "fake" modules that allocate model parameter tensors to the same memory address. The primary objective is to have these modules forward iteratively, paving the way for the generation of multiple accumgrad backward functions anchored to the interchanged weights. This design facilitates the rotation of weights/gradients both before and after gradient generation.

As the tensor transitions to the Flyweight model network layer, it is channeled through multiple module instances that point to the same memory address. This ensures that all operations performed on the tensor are meticulously recorded. During reverse transmission, these operations are replayed, ensuring consistency and accuracy.

In essence, RTP's approach to model initialization is both innovative and efficient. By modifying the model and distributing it across GPUs, we ensure that each device retains only one shard from the onset of the forward process to the conclusion of the backward process. This design not only optimizes memory usage but also streamlines the training process.

\subsection{Partition Strategies}

RTP introduces the concept of a Partition factor, denoted as $N$. This factor represents the number of ranks over which parameters are distributed or sharded. In the context of RTP, activations, weights, and gradients are partitioned by a factor of 
$N$. While activations are partitioned on the batch dimension, similar to strategies in Data Parallelism (DP) and FSDP, weights and gradients undergo custom partitioning tailored to the specific type of layers they belong to three types of Partitioning: Output-Partition, Number-of-head-Partition and Expert Partition

\begin{itemize}
\item {\bf Output-Partition}: Layers such as Linear, Embedding, and Convolution are partitioned based on the output feature dimension.
Post the RTP-based layer, a concatenation layer is introduced to merge the parallel tensor outputs.

\item {\bf Number-of-head-Partition}: The Multihead attention layer \cite{vaswani2017attention} undergoes a unique partitioning strategy. The linear projections for Q, K, and V calculations are divided into multiple linear projections in a column-parallel manner. This is achieved by adjusting the first dimension, determining how many attentions are computed collectively.
The subsequent Feed Forward Network is partitioned on the input dimension. This is because the preceding multihead layer produces only 1/N of the output. By summing up the N subcomponents of the attention output, we can seamlessly obtain the complete attention output.

\item {\bf Expert Partition}: The MoE (Mixture of Experts) layer \cite{shazeer2017outrageously} is inherently designed to support RTP's partitioning strategy. The traditional all-to-all output communication is replaced by RTP's weight rotating mechanism. This ensures that expert outputs are naturally gathered without the need for extensive communication overhead. In fact, MOE with RTP would improve overall performance againest DP and FSDP.
\end{itemize}

Moreover, to enhance communication efficiency, RTP organizes all parameters within a layer unit post-partitioning into a structure called FlatParameter. This structure amalgamates the communication of its individual parameters and distributes them uniformly across ranks. Conceptually, the FlatParameter is a one-dimensional tensor, crafted by concatenating flattened original parameters and adding padding to the clockwise. For instance, in the context of a linear layer, both the weight and bias are merged. Similarly, for a multi-head attention mechanism, the input and output projections are combined under the FlatParameter umbrella.

\begin{figure}
    \centering
  \includegraphics[width=0.85\linewidth]{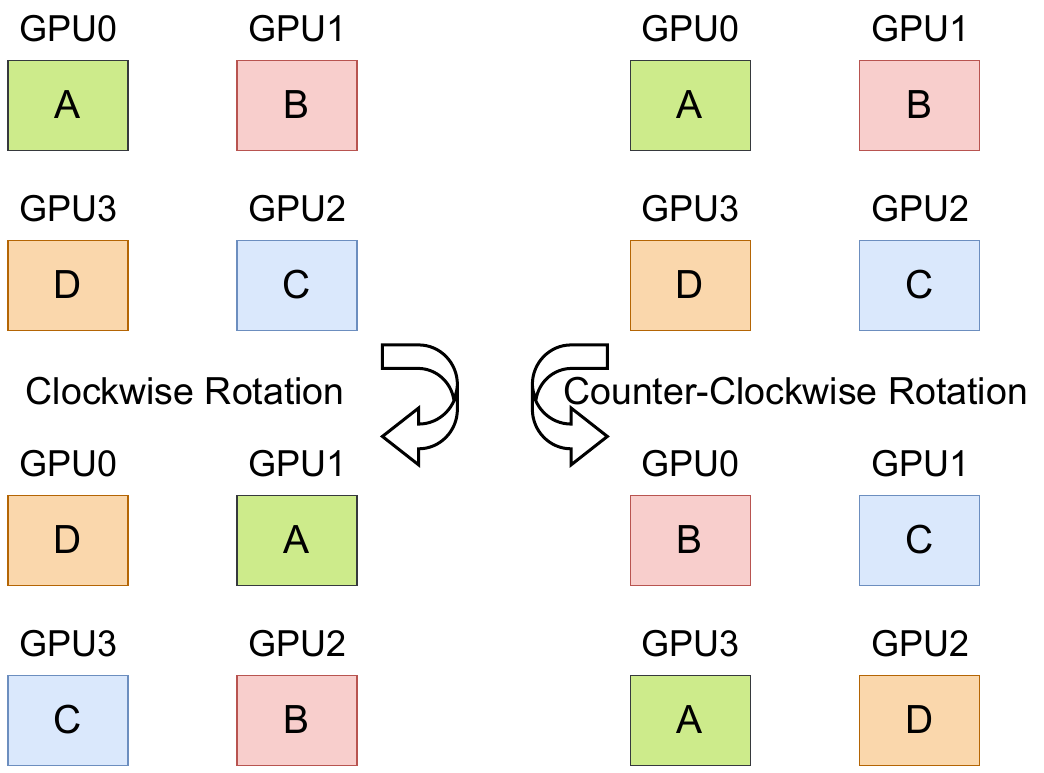}
  \caption{Clockwise Rotation and Counter-clockwise Rotation Primitives Across 4 GPUs }
  \label{fig:Rotations}
\end{figure}


\begin{figure*}[!t]
    \centering
    \begin{subfigure}
    \centering
        \includegraphics[scale=0.375]{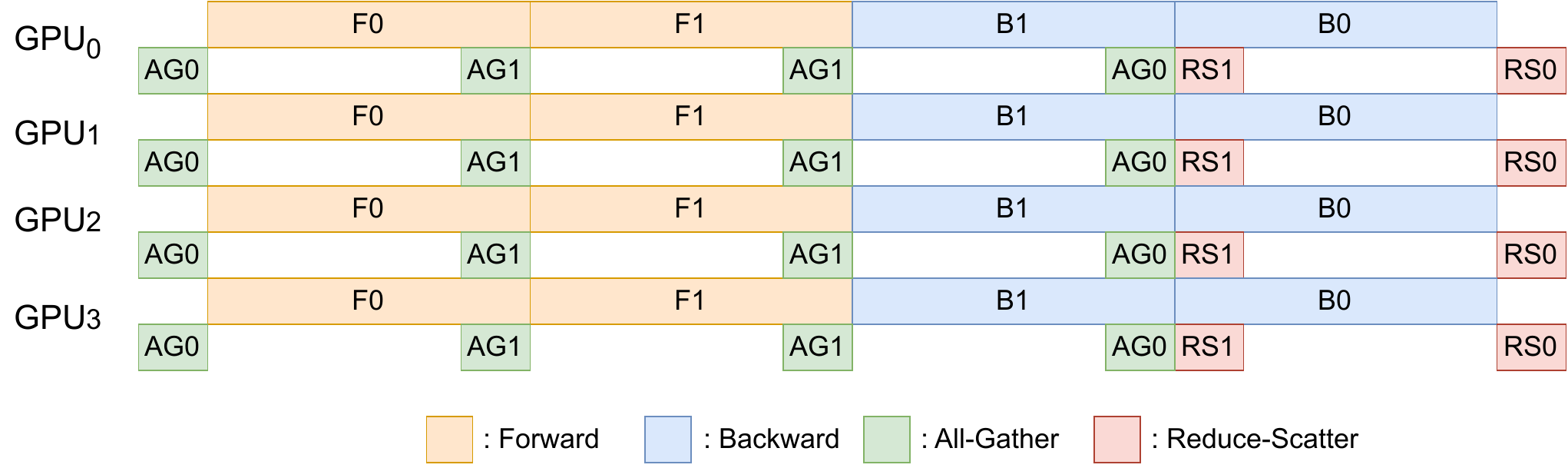}
        \vspace{-0.1in}
        \caption{FSDP Parallelism}
        \vspace{0.15in}
        \label{fig:schedule_breadth_first}
    \end{subfigure}
    \begin{subfigure}
    \centering
        \includegraphics[scale=0.375]{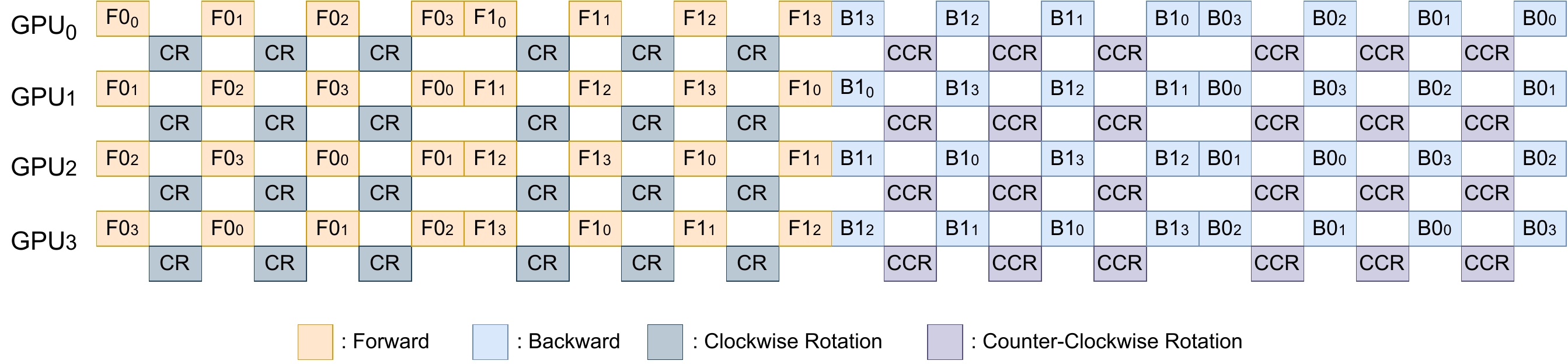}
        \vspace{-0.15in}
        \caption{RTP-inplace Parallelism}
        \vspace{0.15in}
        \label{fig:RTP_flow2}
    \end{subfigure}
    \begin{subfigure}
    \centering
        \includegraphics[scale=0.375]{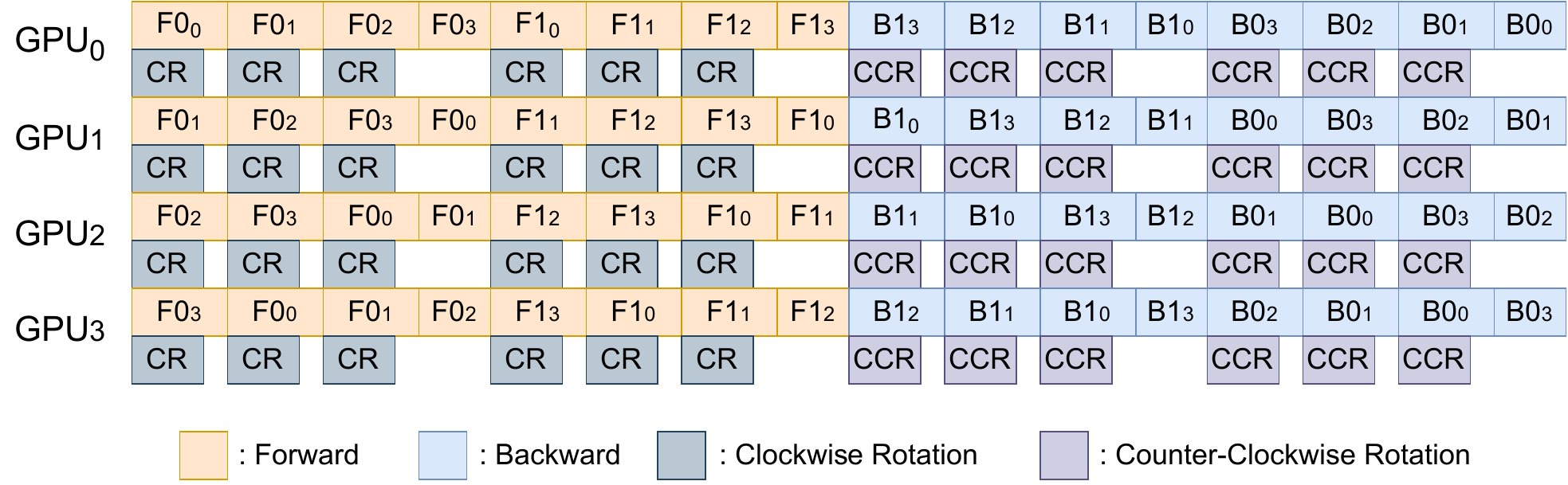}
        \vspace{-0.1in}
        \caption{RTP-out-of-place Parallelism}
        \label{fig:RTP_flow}
    \end{subfigure}
\end{figure*}

\subsection{Rotating Communication Primative}
RTP employs two foundational techniques for communication optimization: Clockwise Rotation and Counter-clockwise Rotation as shown in figure \ref{fig:Rotations}. These techniques are inspired by the ring allreduce concept \cite{you2018imagenet}, where each worker sends and receives distinct messages to and from its adjacent workers. Specifically, we employ clockwise rotation for the forward communication of weight and counter-clockwise rotation for the backward communication of weight and gradient. Here the rotation communication primative is mainly focusing on prefetching the next weights  before the next partition forward  operator and prefetching the previous weights and gradients before the next partition forward  operator.

The primary challenge in optimizing communication is the dominance of latencies in small transfers. As the scale grows, both the Partition Weight and gradient size diminish. Although the previously proposed FlatParameter offers some relief, it doesn't provide a comprehensive solution. To address the aforementioned challenges, we propose two primary strategies:
\begin{itemize}
\item {\bf Out-of-place rotation}: This strategy emphasizes overlapping, allowing for simultaneous communication and computation. By leveraging multiple CUDA streams, we achieve this overlap. However, this approach necessitates additional buffer storage. The Pytorch distributed\_c10d library offers a batch\_isend\_irecv abstraction, representing a set of Send/Recv primitives that can be used concurrently. Furthermore, the NCCL backend implementation in Pytorch provides an internal NCCL stream for each device. This stream is used for asynchronous execution, typically on the default stream for calculations. These asynchronous collectives return Work objects, and invoking Work.wait() blocks the CPU thread until the collective completes. To achieve complete overlapping, RTP employs separate CUDA streams for clockwise and counter-clockwise rotation operations, ensuring the overlap of these calculations.

\item {\bf In-place rotation}: This strategy focuses on blocking both communication and computation. Training models with multiple streams or processes often leads to increased memory usage due to parallel data loading and the accumulation of intermediate variables. The challenge of memory deduplication arises when using out-of-place rotation with overlapping. To address this, we propose in-place rotation with a non-overlapping pattern, effectively eliminating memory deduplication. This approach, devoid of the need for additional buffers, aligns closely with the memory cost of idealized computer.
\end{itemize}

\subsection{Analysis}
\label{subsec:bf_theory}

\subsubsection{Computational Efficiency}

\label{sec:Computational Efficiency}
\begin{equation}
    E_{compute}= N \times Kernel(\frac{B}{N}, I, \frac{O}{N}).
    \label{eq: Computational Efficiency}
\end{equation}
The theoretical underpinning of RTP's computational efficiency posits it as tantamount to $\frac{1}{N}$ local computation,  and tantamount to DP/TP/FSDP computation. Yet, practical GPU computations, especially when handling diminutive kernels, are profoundly influenced by the kernel's size and architectural configuration. Specifically, segmenting the weight into N distinct portions and executing it iteratively N times tends to be suboptimal compared to a singular execution of the entire kernel. This inefficiency is predominantly attributed to:

Kernel Launch Overheads: The cumulative overheads introduced by multiple kernel launches can be detrimental, especially when executed in succession.

GPU Occupancy Concerns: Minuscule kernels often harness only a fraction of the GPU's computational prowess. This underutilization can lead to significant portions of the GPU remaining dormant.

To circumvent these challenges, one potential strategy is to augment the kernel size. This can potentially attenuate the non-linear slowdowns inherent with smaller kernels. Further empirical evaluations, as discussed in the experimental section, elucidate that RTP's computational efficiency, especially for expansive language models, can be enhanced to approximate 90\% of the FSDP implementation.

\subsubsection{Communication Efficiency}
\begin{equation}
    E_{communicate}= (N - 1)\times Send/Recv(M/N).
    \label{eq: Computational Efficiency}
\end{equation}

From a theoretical vantage point, RTP's communication efficiency mirrors that of FSDP, particularly in the context of the allgather operation's communication cost, which entails the aggregation of 1/N weight fraction iteratively for (N-1) times. Empirical evaluations, specifically our custom NCCL-test executed on 8 GPUs, corroborate that the temporal cost associated with the clockwise-rotation and counter-clockwise-rotation communication operations exhibits a near-linear relationship with the allgather operation, especially when the message size surpasses 1MB.

\begin{figure*}
    \centering
  \includegraphics[scale=0.37]{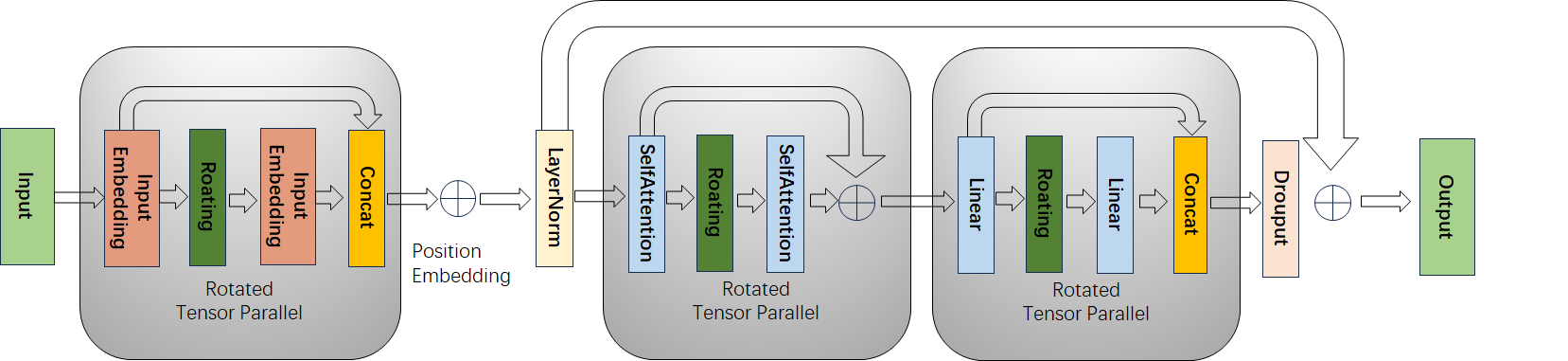}
  \caption{foundation model with Rotated Tensor Paralleism, rotating is clockwise rotating in the forward pass and counter-clockwise roating in the backward pass }
  \label{fig:Treansfermer model with RTP}
\end{figure*}

\begin{figure*}
    \centering
  \includegraphics[scale=0.37]{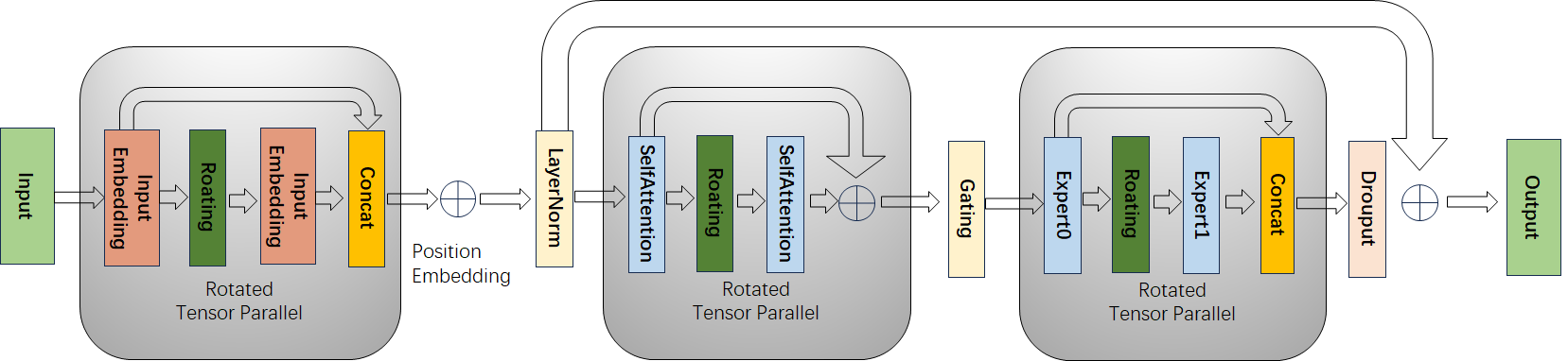}
  \caption{MOE-based foundation model with Rotated Tensor Paralleisn, rotating is clockwise rotating in the forward pass and counter-clockwise roating in the backward pass }
  \label{fig:MOE model with RTP}
    \vspace{-0.15in}
\end{figure*}

\subsubsection{Overlapping}

In the domain of distributed training, the concurrent execution of communication and computation tasks is of paramount importance to maximize efficiency. RTP  ensure that GPU devices are consistently and fully utilized throughout the distributed training process, thereby minimizing potential downtime attributed to non-computational operations. This is achieved by retaining computationally capable shards in the distribution, which allows computation and communication to start simultaneously, whether forward or backward propagation

Figure \ref{fig:RTP_flow} presents a granular understanding of RTP's overlapping capabilities. This illustrative representation provides insights into the concurrent execution of tasks in RTP, emphasizing its efficiency. Compared with FSDP.
A salient feature of RTP is its expedited startup time which is architected to initiate both computation and communication synchronously. This concurrent initiation not only streamlines the training process but also ensures a swifter startup time for RTP, positioning it as a superior alternative in terms of efficiency.

RTP out-of-place(figure \ref{fig:RTP_flow}) additionally achieves efficient statistical calculation overlap by building communication storage space, while RTP-inplace (figure \ref{fig:RTP_flow2})does not require additional space and is closer to ideal storage. These two methods embody the tradoff of computational efficiency and storage efficiency.

\subsubsection{Memory Arrangement}
The efficacy of distributed training paradigms is often contingent upon judicious memory management. Our RTP framework introduces a nuanced approach to memory arrangement, optimizing both space and computational efficiency. 
RTP out-of-place earmarks a distinct communication buffer exclusively for rotation operations. This strategic allocation not only streamlines the rotation process but also engenders efficient overlapping, a critical facet for enhancing throughput in distributed training.

Within the RTP out-of-place paradigm shown in \ref{fig:RTP_flow}, rotation operations recur N-1 times, parallel against computation operations that iterate N times. This differential come from local copy of the initital weight. 
Our memory arrange give the pre-defined TTL of the communication buffer, which invariably terminates prior to the culmination of the last iteration, there emerges an opportunity for memory recycle. Specifically, the  memory segment, once dedicated to the communication buffer, is adeptly repurposed for engendering output activations.
This dynamic memory reallocation mechanism is emblematic of RTP's commitment to optimal resource utilization. In scenarios where activation memory requisites either mirror or surpass those of weights and gradients, RTP's out-of-place strategy exhibits a memory footprint that approximates the theoretical optima.

\section{RTP Transformers}

We take advantage of the structure of transformer networks to create a simple RTP implementation with our customized rotation primitives and a few merge operation. A transformer layer consists
of a self attention block followed by a feed forward layer.
perceptron (MLP) as shown in Figure \ref{fig:Treansfermer model with RTP}. We introduce
RTP in Embedding, linear and Self-attention. A variant  transformer layer replace feed forward layer with mix-of-expert feed forward layer.

\begin{itemize}
\item {\bf Embedding and Linear Block}:
Both the embedding and linear layers in transformers can be represented using General Matrix Multiply (GEMM) operations. Drawing inspiration from Megatron-LM, the weight matrix is split along its columns, which corresponds to the output dimension. The initial segment of the MLP block involves a GEMM operation succeeded by a GeLU nonlinearity. This design mandates N GEMM operations coupled with N rotation communications. The forward pass concludes with a concatenation operation, while the backward pass remains unaffected.

\begin{equation}
    [Y_1, Y_2] = [\begin{bmatrix}
        GEMM(X_1, A_1) \\
        GEMM(X_2, A_1)
    \end{bmatrix}, \begin{bmatrix}
        GEMM(X_1, A_2) \\
        GEMM(X_2, A_2)
    \end{bmatrix}]
    \label{eq: Computational Efficiency}
\end{equation}

\item {\bf Multiple-head-Attention Block}:
The multihead attention operation is inherently parallelizable. The GEMMs associated with the key (K), query (Q), and value (V) matrices are partitioned in a column-parallel manner. This design choice ensures that each attention head's matrix multiplication is localized to a single GPU. The subsequent GEMM, stemming from the output linear layer post self-attention, is parallelized along its rows, corresponding to the input dimension. This architecture necessitates N attention operations paired with N rotation communications. The forward pass culminates in an addition operation, leaving the backward pass unaltered.

\begin{equation}
\begin{split}
    [Y_1, Y_2] = [
        &ATTN(X_1, A_1) + 
        ATTN(X_2, A_1)
, \\
        &ATTN(X_1, A_2) +
        ATTN(X_2, A_2)
]
    \label{eq: Computational Efficiency}
\end{split}
\end{equation}

\item {\bf MOE Block}:
The Mixture of Experts (MOE) block introduces a new dimension to the RTP framework. Traditional Data Parallelism (DP) and Fully Sharded Data Parallelism (FSDP) necessitate the insertion of all-to-all operations both before and after MOE computations. However, RTP, with its rotation mechanism, offers a more streamlined approach as shown in figure \ref{fig:MOE model with RTP}. Expert rotation within the MOE block is executed as a sequential flow: starting with expert0, followed by a rotation operation, transitioning to expert1, and culminating in a concatenation operation. Furthermore, the RTP-inplace variant allows for concurrent communication of expert weights during MOE computations, enhancing computational efficiency. This integration of expert computations and communications within the RTP framework ensures enhanced performance and reduced overheads.

\end{itemize}

Our RTP approach can be characterized as techniques aimed at reducing memory duplication. We present further details about the RTP model with other rotation generation in Appendix for reference. In summary, our approach as described above is simple to implement, requiring only a few extra contact and add operations added to the forward and backward pass. It is orthogonal and complementary to the pipeline model parallelism.

\section{Evaluations}

\begin{table*}
    \centering
    \begin{tabular}{c|c|c|c|c|c|c}
        Models & Attention &  Hidden &  Layers & sequence &Vocab & Embedding\\
         & Heads & Size &  & length & Size & Size\\
        \hline
        GPT2 (117M) & 16 & 768 & 12 & 512 & 50257 & 3072\\
        BERT-large (340M) & 16 & 1024 & 24 & 512 & 30522 & 4096\\
        GPT2* (500M)  & 16 & 1280 & 20 & 1024 & 50257 & 5120\\
        GPT2-large (774M)  & 16 & 1280 & 32 & 1024 & 50257 & 5120\\
        GPT2-XL (1.5B)  & 16 & 1600 & 48 & 1024 & 50257 & 6400\\
        GPT2-neo (2.7B)  & 16 & 2560 & 32 & 1024 & 50257 & 10240\\
        \hline
    \end{tabular}
    \caption{Model configurations used during evaluation. GPT2 (500M) is our customized model which can fit into A100 80 GPU with 8 batchsize}
    \vspace{-0.15in}
    \label{tab:configuration}
\end{table*}

We embarked on a rigorous evaluation of RTP on state-of-the-art neural network models, contrasting the outcomes with those of other prevalent techniques. The specifics of the experiment are delineated in Section \ref{sec:Experiment Setup}. The experiments are categorized into three distinct sections. Section \ref{sec:Evaluation of R-TP's Memory Efficiency} delves into RTP's proficiency in handling models of varying magnitudes. Section \ref{sec:Training Speed} sheds light on the implications of memory deduplication on training dynamics.

\subsection{Experiment Setup}
\label{sec:Experiment Setup}

For these evaluations, we utilized models such as GPT2, BERT-large, GPT2-large, GPT2-XL and GPT2-neo. The LLM models were trained using RTP across 8 A100 80GB GPUs, interconnected by a NVILINK. The primary aim was to gauge the memory efficacy RTP in both training and inference of large-scale neural network models. Furthermore, we deployed these networks to assess the performance with the maximum batch size available. Metrics employed in these experiments included Throughput per GPU, peak memory allocated.

\subsection{Memory Efficiency}
\label{sec:Evaluation of R-TP's Memory Efficiency}

In this section, we delve into the performance of RTP when handling models of diverse scales, ranging from smaller configurations to the massive GPT-XL 1.5B, and juxtapose the results with FSDP and DDP. These configuration is shown on \ref{tab:configuration}, where the batchsize is set as 1.

\begin{figure}[t]
    \centering
    \includegraphics[width=\linewidth]{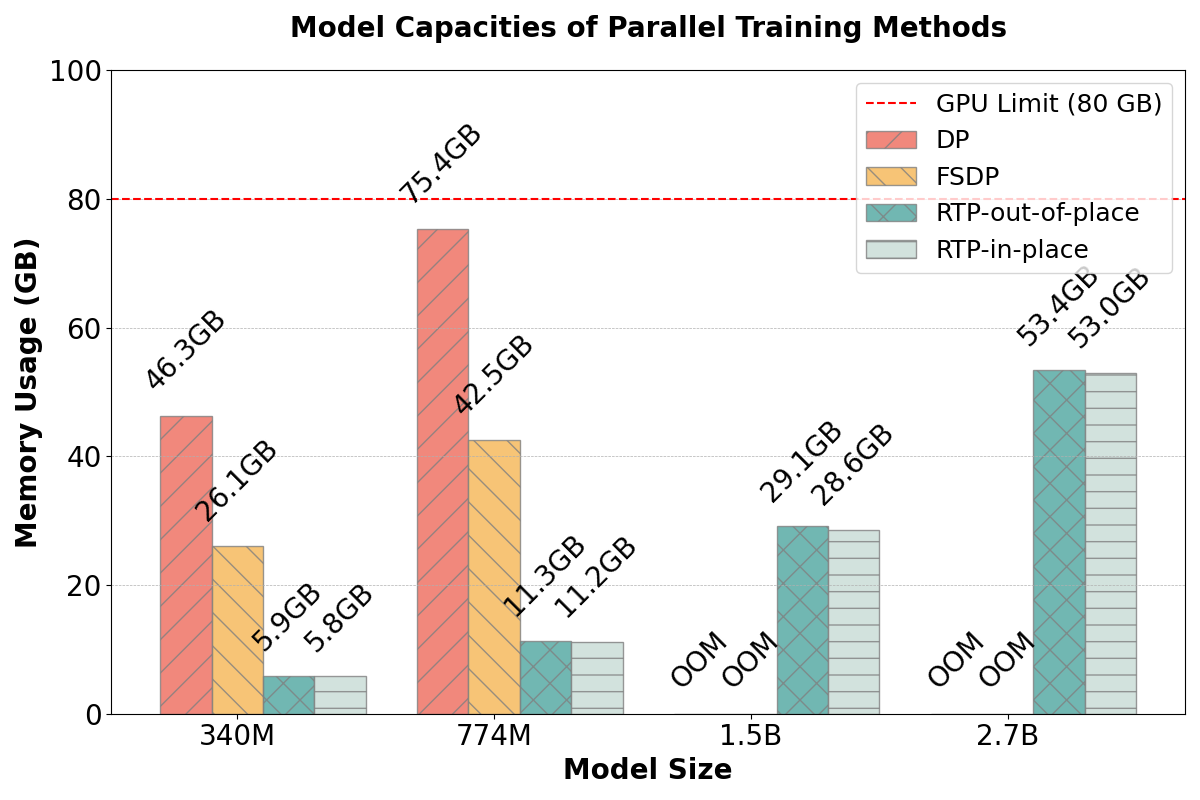}
    \caption{Model Capacity Evaluation. Tested with a DGX-A100 station with 8 A100 (80GB) GPUs. All the \textsc{local\_batch\_size} is 1.}
    \label{fig:model_capacity}
\end{figure}

The experimental outcomes are illustrated in Figure \ref{fig:model_capacity}. The memory efficiency of RTP gains over 75\% that of FSDP and 85\% of DP when evaluating models up to the scale of 774M. However, while FSDP stop with memory constraints for models surpassing this threshold, RTP seamlessly accommodates the GPT-XL model (1.5B), showcasing its prowess in memory efficiency and optimization. 

  
  

\begin{figure}[t]
    \centering
    \includegraphics[width=\linewidth]{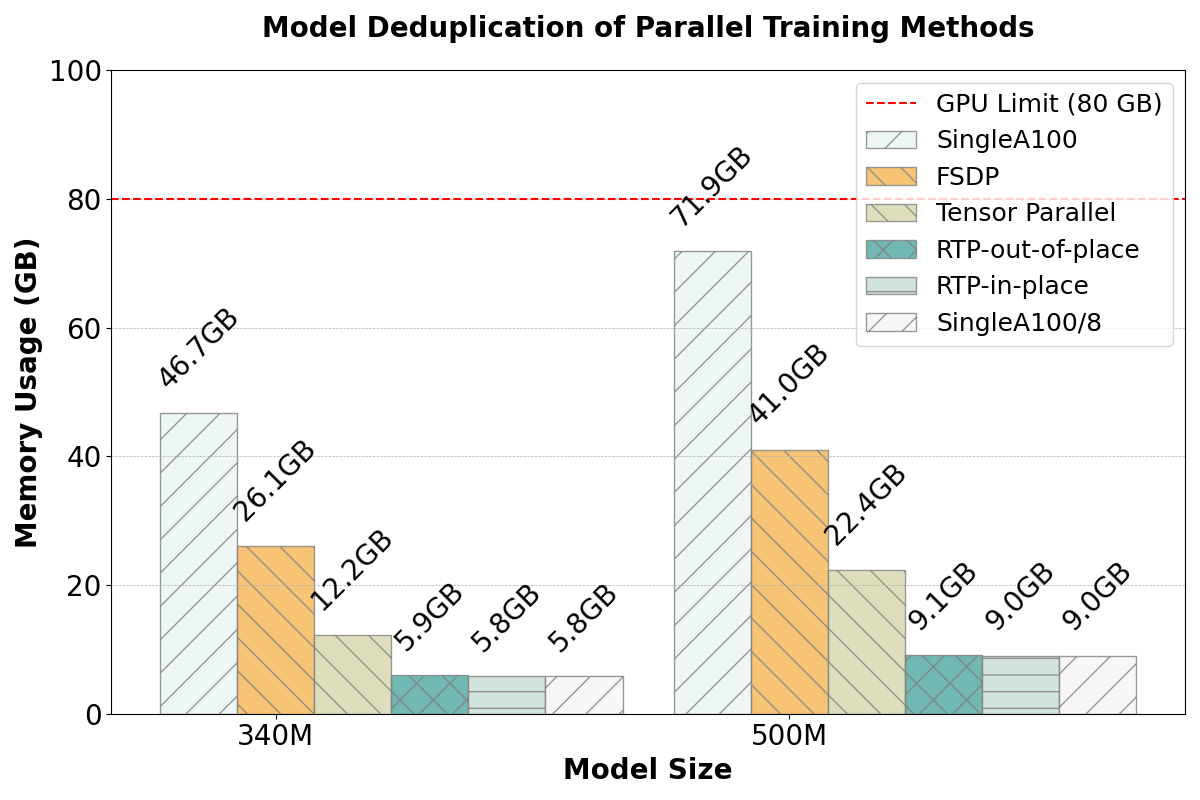}
  
    \vspace{-0.15in}
  \caption{Model Deduplication Evaluation. Tested with a DGX-A100 station with 8 A100 (80GB) GPUs. All the \textsc{global\_batch\_size} is 8.}
  
  \label{fig:Memory Deduplication}
\end{figure}

\begin{figure}[t]
    \centering
  \includegraphics[width=\linewidth]{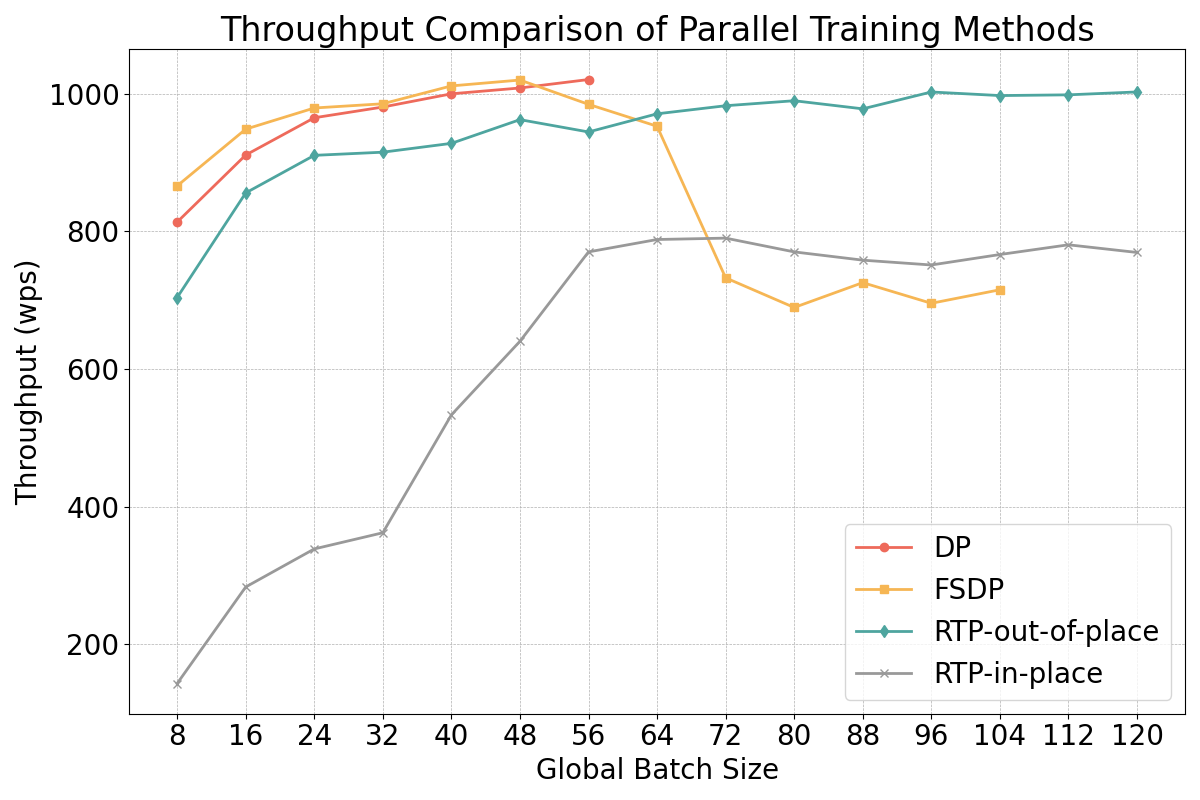}
    \vspace{-0.15in}
  \caption{Throughput evaluation for GPT2-500M}
  
    \vspace{-0.15in}
  \label{fig:Throughput evaluation for GPT-2}
\end{figure}

\begin{figure}[t]
    \centering
  \includegraphics[width=\linewidth]{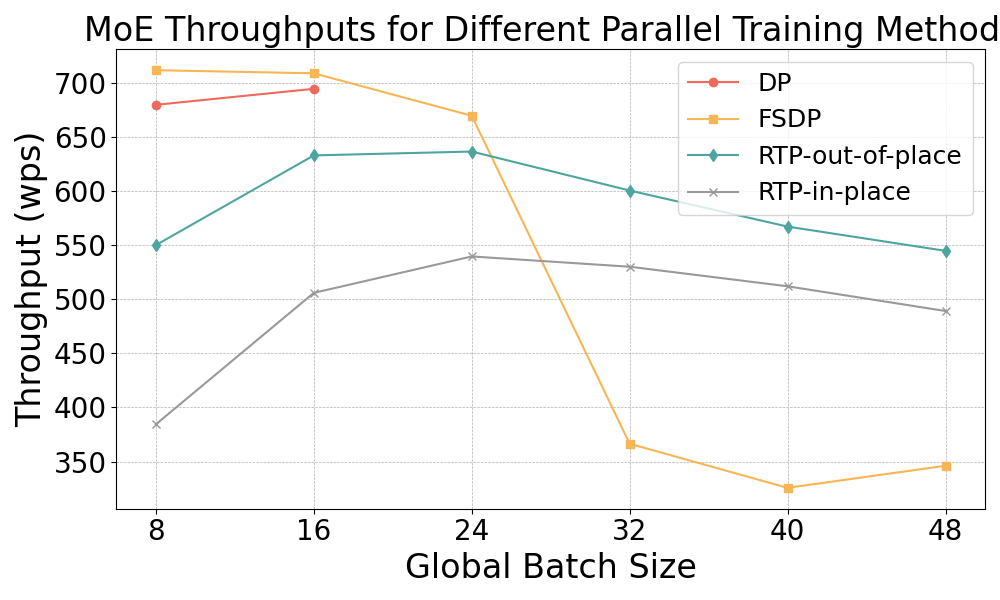}
    \vspace{-0.15in}
  \caption{Throughput evaluation for MOE GPT2-500M}
  
    \vspace{-0.15in}
  \label{fig:Throughput evaluation for GPT-2}
\end{figure}


\subsection{Memory Deduplication}
To empirically validate the memory deduplication prowess of our proposed algorithm, we embarked on a comprehensive evaluation involving three distinct neural network architectures: GPT, BERT-large, and a GPT-up-to-A100 tailored to closely match the specifications of the A100 80G. These evaluations were conducted on the single machine with single-device DDP as idealized computer, a benchmark for assessing algorithmic efficiency in real-world scenarios.

For the experiment, we set the global batch size to 8, implying that on a singular A100 card, eight distinct sets of distributed algorithms were executed concurrently on the GPU, with each set having a batch size per GPU (global-batch-size/gpu) of 1. The memory storage overhead on each card was then meticulously recorded, and subsequently multiplied by eight to facilitate a direct comparison with the storage overhead observed on the single GPU.

The figure \ref{fig:Memory Deduplication} offers some compelling insights. Both RTP-inplace and RTP-outplace variants showcased memory storage metrics that were in close alignment with those of the single machine. This is a testament to the efficiency of RTP in optimizing memory usage without compromising on performance. In stark contrast, both FSDP and TP exhibited storage overheads that were two to four times greater than the theoretical benchmarks set by the single machine.

This disparity underscores the optimized memory deduplication capabilities of RTP. The results unequivocally affirm that RTP, both in its inplace and outplace variants, is adept at ensuring optimal memory usage, making it a formidable tool for researchers and practitioners aiming to maximize computational efficiency while minimizing memory overheads.

\subsection{Training Speed}
\label{sec:Training Speed}

To assess the overall efficiency of our approach, we conducted an evaluation of the end-to-end iteration time for the GPT-up-to-A100, as depicted in figure \ref{fig:Throughput evaluation for GPT-2}. Our empirical results demonstrate that the RTP, as introduced in this study, results in a throughput reduction ranging from -13\% to -1.7\% when compared to data parallelism. Furthermore, when juxtaposed against FSDP, the performance fluctuates between a reduction of -10\% to -1.6\%. Notably, as the batch size augments in tandem with the kernel size, the performance disparity narrows, eventually converging at approximately 1000 wps. The intricate relationship between RTP performance, kernel size, and batch size augmentation is elaborated upon in section \ref{sec:Computational Efficiency}. A particularly interesting thing is that when the batchsize is full, the FSDP throughput drops sharply and is strictly weaker than RTP by more than 50

In addition to the aforementioned evaluations, we also turned our analytical lens towards the Mixture of Experts (MoE) paradigm. MoE, renowned for its ability to efficiently route input data to specialized sub-networks, was evaluated in tandem with the throughput performances of various methodologies, including DP, FSDP, and both RTP variants. Our observations indicate that RTP would cause throughput reduction ranging from -23\% to -10\% compared to data parallelism and -19\% to -9\%. Also the large batchsize downgrade case is also shown in MOE evaluations.

An interesting thing is that when the batchsize is increased, the additional overhead on FSDP storage is slowly smoothed out, which allows RTP and FSDP to reach the close maximum batchsize setting. This can be considered as the Memory Duplication of FSDP will be recycled and acted upon. This effectively eliminates additional storage overhead due to reuse, which would be discussed on appendix \ref{sec:Peak Memory Scale with batch}.
But at the same time, after the batchsize is increased, Throughput of RTP can outperform FSDP. We attribute this to the perfect overlapping of RTP. There is no additional waiting time for calculation and communication in RTP, but FSDP needs to wait for the first allgather to start working. This Brings additional throughput benefits to RTP.

Moreover, RTP can achieve 10\% to 40\% performance gain on V100 GPU with PCIE, refer appendix \ref{sec:V100 Throughput} for more information.

\section{Conclusion}
The rapid evolution of neural network models has necessitated innovative solutions to address the challenges of scalability and efficiency in training. The paper introduces Rotated Tensor Parallelism (RTP), a groundbreaking approach that focuses on memory deduplication in distributed training environments. By strategically decomposing and sharding batch tensors, RTP optimizes memory consumption by emphasizing both activation and parameter deduplication. When compared with existing parallelism techniques, RTP stands out in its ability to reduce memory overheads and enhance training performance. Notably, RTP aligns closely with the optimal memory benchmarks set by the unlimited memory idealized computer and offers significant memory savings compared to prevalent methods like Fully Sharded Data Parallelism (FSDP). The introduction of RTP underscores the importance of memory deduplication in enhancing the efficiency of distributed training systems and positions it as a promising alternative to existing paradigms.

\section*{Acknowledgements}
This work is partially supported by DE-SC0023452 grant from the Department of Energy. The authors are thankful for their support.

\nocite{langley00}

\bibliography{example_paper}

\begin{thebibliography}{29}
\providecommand{\natexlab}[1]{#1}
\providecommand{\url}[1]{\texttt{#1}}
\expandafter\ifx\csname urlstyle\endcsname\relax
  \providecommand{\doi}[1]{doi: #1}\else
  \providecommand{\doi}{doi: \begingroup \urlstyle{rm}\Url}\fi

\bibitem[In-()]{In-place_Operations}
{In-place Operations}.
\newblock \url{https://docs.nvidia.com/deeplearning/nccl/user-guide/docs/usage/inplace.html}.

\bibitem[a10()]{a100_80G}
{NVIDIA A100 Tensor Core GPU}.
\newblock \url{https://www.nvidia.com/en-us/data-center/a100/}.

\bibitem[Bugnion et~al.(1997)Bugnion, Devine, Govil, and Rosenblum]{bugnion1997disco}
Bugnion, E., Devine, S., Govil, K., and Rosenblum, M.
\newblock Disco: Running commodity operating systems on scalable multiprocessors.
\newblock \emph{ACM Transactions on Computer Systems (TOCS)}, 15\penalty0 (4):\penalty0 412--447, 1997.

\bibitem[Dao et~al.(2022)Dao, Fu, Ermon, Rudra, and R{\'e}]{dao2022flashattention}
Dao, T., Fu, D., Ermon, S., Rudra, A., and R{\'e}, C.
\newblock Flashattention: Fast and memory-efficient exact attention with io-awareness.
\newblock \emph{Advances in Neural Information Processing Systems}, 35:\penalty0 16344--16359, 2022.

\bibitem[Dean et~al.(2012)Dean, Corrado, Monga, Chen, Devin, Mao, Ranzato, Senior, Tucker, Yang, et~al.]{dean2012large}
Dean, J., Corrado, G., Monga, R., Chen, K., Devin, M., Mao, M., Ranzato, M., Senior, A., Tucker, P., Yang, K., et~al.
\newblock Large scale distributed deep networks.
\newblock \emph{Advances in neural information processing systems}, 25, 2012.

\bibitem[Dehghani et~al.(2023)Dehghani, Djolonga, Mustafa, Padlewski, Heek, Gilmer, Steiner, Caron, Geirhos, Alabdulmohsin, et~al.]{dehghani2023scaling}
Dehghani, M., Djolonga, J., Mustafa, B., Padlewski, P., Heek, J., Gilmer, J., Steiner, A.~P., Caron, M., Geirhos, R., Alabdulmohsin, I., et~al.
\newblock Scaling vision transformers to 22 billion parameters.
\newblock In \emph{International Conference on Machine Learning}, pp.\  7480--7512. PMLR, 2023.

\bibitem[Harmes \& Diaz(2008)Harmes and Diaz]{harmes2008flyweight}
Harmes, R. and Diaz, D.
\newblock The flyweight pattern.
\newblock \emph{Pro JavaScript Design Patterns}, pp.\  179--195, 2008.

\bibitem[He et~al.(2021)He, Li, Soltanolkotabi, and Avestimehr]{pipetransformer}
He, C., Li, S., Soltanolkotabi, M., and Avestimehr, S.
\newblock Pipetransformer: Automated elastic pipelining for distributed training of transformers.
\newblock \emph{arXiv preprint arXiv:2102.03161}, 2021.

\bibitem[Huang et~al.(2019)Huang, Cheng, Bapna, Firat, Chen, Chen, Lee, Ngiam, Le, Wu, et~al.]{gpipe}
Huang, Y., Cheng, Y., Bapna, A., Firat, O., Chen, D., Chen, M., Lee, H., Ngiam, J., Le, Q.~V., Wu, Y., et~al.
\newblock Gpipe: Efficient training of giant neural networks using pipeline parallelism.
\newblock \emph{Advances in neural information processing systems}, 32, 2019.

\bibitem[Imambi et~al.(2021)Imambi, Prakash, and Kanagachidambaresan]{imambi2021pytorch}
Imambi, S., Prakash, K.~B., and Kanagachidambaresan, G.
\newblock Pytorch.
\newblock \emph{Programming with TensorFlow: Solution for Edge Computing Applications}, pp.\  87--104, 2021.

\bibitem[Kim et~al.(2020)Kim, Lee, Jeong, Baek, Yoon, Kim, Lim, and Kim]{kim2020torchgpipe}
Kim, C., Lee, H., Jeong, M., Baek, W., Yoon, B., Kim, I., Lim, S., and Kim, S.
\newblock torchgpipe: On-the-fly pipeline parallelism for training giant models.
\newblock \emph{arXiv preprint arXiv:2004.09910}, 2020.

\bibitem[Langley(2000)]{langley00}
Langley, P.
\newblock Crafting papers on machine learning.
\newblock In Langley, P. (ed.), \emph{Proceedings of the 17th International Conference on Machine Learning (ICML 2000)}, pp.\  1207--1216, Stanford, CA, 2000. Morgan Kaufmann.

\bibitem[Li et~al.(2020)Li, Zhao, Varma, Salpekar, Noordhuis, Li, Paszke, Smith, Vaughan, Damania, et~al.]{li2020pytorch}
Li, S., Zhao, Y., Varma, R., Salpekar, O., Noordhuis, P., Li, T., Paszke, A., Smith, J., Vaughan, B., Damania, P., et~al.
\newblock Pytorch distributed: Experiences on accelerating data parallel training.
\newblock \emph{arXiv preprint arXiv:2006.15704}, 2020.

\bibitem[Liberty et~al.(2020)Liberty, Karnin, Xiang, Rouesnel, Coskun, Nallapati, Delgado, Sadoughi, Astashonok, Das, et~al.]{liberty2020elastic}
Liberty, E., Karnin, Z., Xiang, B., Rouesnel, L., Coskun, B., Nallapati, R., Delgado, J., Sadoughi, A., Astashonok, Y., Das, P., et~al.
\newblock Elastic machine learning algorithms in amazon sagemaker.
\newblock In \emph{Proceedings of the 2020 ACM SIGMOD International Conference on Management of Data}, pp.\  731--737, 2020.

\bibitem[Narayanan et~al.(2019)Narayanan, Harlap, Phanishayee, Seshadri, Devanur, Ganger, Gibbons, and Zaharia]{narayanan2019pipedream}
Narayanan, D., Harlap, A., Phanishayee, A., Seshadri, V., Devanur, N.~R., Ganger, G.~R., Gibbons, P.~B., and Zaharia, M.
\newblock Pipedream: Generalized pipeline parallelism for dnn training.
\newblock In \emph{Proceedings of the 27th ACM Symposium on Operating Systems Principles}, pp.\  1--15, 2019.

\bibitem[Narayanan et~al.(2021)Narayanan, Shoeybi, Casper, LeGresley, Patwary, Korthikanti, Vainbrand, Kashinkunti, Bernauer, Catanzaro, et~al.]{narayanan2021efficient}
Narayanan, D., Shoeybi, M., Casper, J., LeGresley, P., Patwary, M., Korthikanti, V., Vainbrand, D., Kashinkunti, P., Bernauer, J., Catanzaro, B., et~al.
\newblock Efficient large-scale language model training on gpu clusters using megatron-lm.
\newblock In \emph{Proceedings of the International Conference for High Performance Computing, Networking, Storage and Analysis}, pp.\  1--15, 2021.

\bibitem[OpenAI(2023)]{openai2023gpt4}
OpenAI.
\newblock Gpt-4 technical report, 2023.

\bibitem[Pope et~al.(2023)Pope, Douglas, Chowdhery, Devlin, Bradbury, Heek, Xiao, Agrawal, and Dean]{pope2023efficiently}
Pope, R., Douglas, S., Chowdhery, A., Devlin, J., Bradbury, J., Heek, J., Xiao, K., Agrawal, S., and Dean, J.
\newblock Efficiently scaling transformer inference.
\newblock \emph{Proceedings of Machine Learning and Systems}, 5, 2023.

\bibitem[Ren et~al.(2021)Ren, Rajbhandari, Aminabadi, Ruwase, Yang, Zhang, Li, and He]{ren2021zero}
Ren, J., Rajbhandari, S., Aminabadi, R.~Y., Ruwase, O., Yang, S., Zhang, M., Li, D., and He, Y.
\newblock $\{$ZeRO-Offload$\}$: Democratizing $\{$Billion-Scale$\}$ model training.
\newblock In \emph{2021 USENIX Annual Technical Conference (USENIX ATC 21)}, pp.\  551--564, 2021.

\bibitem[Sergeev \& Del~Balso(2018)Sergeev and Del~Balso]{sergeev2018horovod}
Sergeev, A. and Del~Balso, M.
\newblock Horovod: fast and easy distributed deep learning in tensorflow.
\newblock \emph{arXiv preprint arXiv:1802.05799}, 2018.

\bibitem[Shazeer et~al.(2017)Shazeer, Mirhoseini, Maziarz, Davis, Le, Hinton, and Dean]{shazeer2017outrageously}
Shazeer, N., Mirhoseini, A., Maziarz, K., Davis, A., Le, Q., Hinton, G., and Dean, J.
\newblock Outrageously large neural networks: The sparsely-gated mixture-of-experts layer.
\newblock \emph{arXiv preprint arXiv:1701.06538}, 2017.

\bibitem[Shoeybi et~al.(2019)Shoeybi, Patwary, Puri, LeGresley, Casper, and Catanzaro]{shoeybi2019megatron}
Shoeybi, M., Patwary, M., Puri, R., LeGresley, P., Casper, J., and Catanzaro, B.
\newblock Megatron-lm: Training multi-billion parameter language models using model parallelism.
\newblock \emph{arXiv preprint arXiv:1909.08053}, 2019.

\bibitem[Singh et~al.(2020)Singh, Manure, Singh, and Manure]{singh2020introduction}
Singh, P., Manure, A., Singh, P., and Manure, A.
\newblock Introduction to tensorflow 2.0.
\newblock \emph{Learn TensorFlow 2.0: Implement Machine Learning and Deep Learning Models with Python}, pp.\  1--24, 2020.

\bibitem[Sourouri et~al.(2014)Sourouri, Gillberg, Baden, and Cai]{sourouri2014effective}
Sourouri, M., Gillberg, T., Baden, S.~B., and Cai, X.
\newblock Effective multi-gpu communication using multiple cuda streams and threads.
\newblock In \emph{2014 20th IEEE International Conference on Parallel and Distributed Systems (ICPADS)}, pp.\  981--986. IEEE, 2014.

\bibitem[Touvron et~al.(2023)Touvron, Martin, Stone, Albert, Almahairi, Babaei, Bashlykov, Batra, Bhargava, Bhosale, et~al.]{touvron2023llama}
Touvron, H., Martin, L., Stone, K., Albert, P., Almahairi, A., Babaei, Y., Bashlykov, N., Batra, S., Bhargava, P., Bhosale, S., et~al.
\newblock Llama 2: Open foundation and fine-tuned chat models.
\newblock \emph{arXiv preprint arXiv:2307.09288}, 2023.

\bibitem[Vaswani et~al.(2017)Vaswani, Shazeer, Parmar, Uszkoreit, Jones, Gomez, Kaiser, and Polosukhin]{vaswani2017attention}
Vaswani, A., Shazeer, N., Parmar, N., Uszkoreit, J., Jones, L., Gomez, A.~N., Kaiser, {\L}., and Polosukhin, I.
\newblock Attention is all you need.
\newblock \emph{Advances in neural information processing systems}, 30, 2017.

\bibitem[You et~al.(2018)You, Zhang, Hsieh, Demmel, and Keutzer]{you2018imagenet}
You, Y., Zhang, Z., Hsieh, C.-J., Demmel, J., and Keutzer, K.
\newblock Imagenet training in minutes.
\newblock In \emph{Proceedings of the 47th International Conference on Parallel Processing}, pp.\  1--10, 2018.

\bibitem[{\v{Z}}{\'a}k(1983)]{vzak1983turing}
{\v{Z}}{\'a}k, S.
\newblock A turing machine time hierarchy.
\newblock \emph{Theoretical Computer Science}, 26\penalty0 (3):\penalty0 327--333, 1983.

\bibitem[Zhao et~al.(2023)Zhao, Gu, Varma, Luo, Huang, Xu, Wright, Shojanazeri, Ott, Shleifer, et~al.]{zhao2023pytorch}
Zhao, Y., Gu, A., Varma, R., Luo, L., Huang, C.-C., Xu, M., Wright, L., Shojanazeri, H., Ott, M., Shleifer, S., et~al.
\newblock Pytorch fsdp: experiences on scaling fully sharded data parallel.
\newblock \emph{arXiv preprint arXiv:2304.11277}, 2023.

\end{thebibliography}
\bibliographystyle{mlsys2023}

\appendix
\section{Peak Memory Scale with batch}
\label{sec:Peak Memory Scale with batch}

\begin{figure}[h]
    \centering
  \includegraphics[scale=0.65]{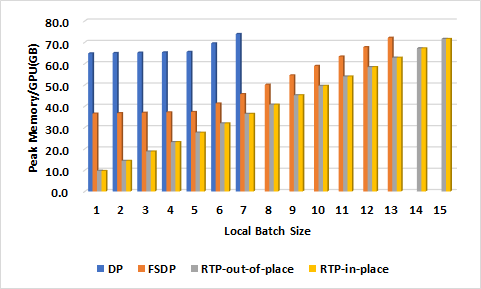}
  
    \vspace{-0.15in}
  \caption{Memory scale with batch size increase}
  
    \vspace{-0.15in}
  \label{fig:Memory scale with batch size increase}
\end{figure}

In our evaluations, as depicted in Figure \ref{fig:Memory scale with batch size increase}, we observed distinct memory scaling behaviors between the Data Parallelism (DP), Fully Sharded Data Parallelism (FSDP), and our proposed RTP methodology. Specifically, while DP and FSDP exhibited a non-linear scaling pattern, RTP demonstrated a linear scaling trend.

A closer examination of the memory arrangement in DP and FSDP reveals that certain memory allocated for weights and gradients can be repurposed for activations. This overlap, while efficient in terms of memory utilization, introduces complexities that can affect the scaling behavior.

An intriguing observation was the behavior of FSDP as the batch size increased. The additional overhead associated with FSDP's memory storage began to diminish, allowing RTP and FSDP to converge towards the similar maximum batch size. This behavior can be attributed to the recycling and subsequent action on the memory duplication in FSDP, effectively nullifying any additional storage overhead through reuse. It has been discussed on FSDP paper's memory arrangement section. 

Put anything that you might normally include after the references as an appendix here, {\it not in a separate supplementary file}. Upload your final camera-ready as a single pdf, including all appendices.


\section{Throughput on V100}
\label{sec:V100 Throughput}

\begin{figure}[t]
    \centering
  \includegraphics[width=\linewidth]{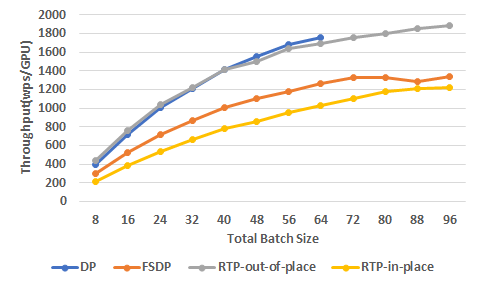}
    \vspace{-0.15in}
  \caption{Throughput evaluation for GPT on V100}
  
    \vspace{-0.15in}
  \label{fig:Throughput evaluation for GPT-2}
\end{figure}

\begin{figure}[t]
    \centering
  \includegraphics[width=\linewidth]{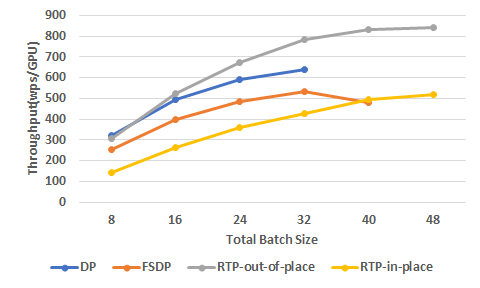}
    \vspace{-0.15in}
  \caption{Throughput evaluation for MOE GPT on V100}
  
    \vspace{-0.15in}
  \label{fig:Throughput evaluation for GPT-2}
\end{figure}

To assess the overall efficiency of our approach, we evaluated the end-to-end iteration time for the GPT variant on
8 V100 32 GB with PCIE connection. Our findings indicate that the RTP presented
in this paper yield a throughput reduce ranging between
21\% and 37.1\% over the data paralleism, and -10\% to
10\% performance reduction/improvement againest FSDP.
The performance gap is becoming smaller when the batch
size increase as kernel size increase.  This enhancement in throughput directly translates to
reduced training durations.

An interesting thing is that when the batchsize is increased,  throughput
of RTP can outperform FSDP on V100. We attribute this to the perfect overlapping of RTP. There is no additional waiting time
for calculation and communication in RTP, but FSDP needs
to wait for the first allgather to start working. This Brings
additional throughput benefits to RTP. A100 with NVILINK bring significant compute and network performace, where the gain of RTP would be minimized and the memory tranfer time would be new challenge. Fourtunaly, it can be solved by coupling with other technologies like flashattention \cite{dao2022flashattention}. This performance improvement is also shown on MOE experiment.

\end{document}